%% file: main.tex
\documentclass[10pt, conference, letterpaper]{IEEEtran}

\usepackage[utf8]{inputenc}
\usepackage{cite}
\usepackage{url}
\usepackage{hyperref}
\usepackage{comment}
\usepackage[normalem]{ulem}
\usepackage{booktabs}
\usepackage{pbox}

\usepackage[pdftex]{graphicx}
\graphicspath{{./fig/}}
\DeclareGraphicsExtensions{.pdf,.jpeg,.png}

\newcommand{\s}[1]{\texttt{#1}}

\usepackage[caption=false,font=footnotesize]{subfig}
\usepackage{fixltx2e}

\usepackage{pgfplots}
\usepackage{todonotes}

\begin{document}

\title{SPIDER: Fault Resilient SDN Pipeline \\ with Recovery Delay Guarantees}

\author{
  \IEEEauthorblockN{
   Carmelo Cascone\IEEEauthorrefmark{1}\IEEEauthorrefmark{2},
   Luca Pollini\IEEEauthorrefmark{3}, 
   Davide Sanvito\IEEEauthorrefmark{3},
   Antonio Capone\IEEEauthorrefmark{1},
   Brunilde Sansò \IEEEauthorrefmark{2}
  }
  \IEEEauthorblockA{
   \IEEEauthorrefmark{1}
   Dipartimento di Elettronica, Informazione e Bioingegneria, Politecnico di Milano, Italy\\
  }
  \IEEEauthorblockA{
   \IEEEauthorrefmark{2}
   D\'epartement de g\'enie \'electrique, \'Ecole Polytechnique de Montr\'eal, Canada\\
  }
  \IEEEauthorblockA{
   \IEEEauthorrefmark{3}
   CNIT, Consorzio Nazionale Interuniversitario per le Telecomunicazioni,  Italy\\
  }
}

\maketitle

\begin{abstract}

When dealing with node or link failures in Software Defined Networking (SDN), the network capability to establish an alternative path depends on controller reachability and on the round trip times (RTTs) between controller and involved switches. Moreover, current SDN data plane abstractions for failure detection (e.g. OpenFlow ``Fast-failover'') do not allow programmers to tweak switches' detection mechanism, thus leaving SDN operators still relying on proprietary management interfaces (when available) to achieve guaranteed detection and recovery delays. We propose SPIDER, an OpenFlow-like pipeline design that provides i) a detection mechanism based on switches' periodic link probing and ii) fast reroute of traffic flows even in case of distant failures, regardless of controller availability. SPIDER can be implemented using stateful data plane abstractions such as OpenState or Open vSwitch, and it offers guaranteed short (i.e. ms) failure detection and recovery delays, with a configurable trade off between overhead and failover responsiveness. We present here the SPIDER pipeline design, behavioral model, and analysis on flow tables' memory impact. We also implemented and experimentally validated SPIDER using OpenState (an OpenFlow 1.3 extension for stateful packet processing), showing numerical results on its performance in terms of recovery latency and packet losses.

\end{abstract}

\section{Introduction}
\label{sec:introduction}

The longly anticipated paradigm shift of Software Defined Networking (SDN) is radically transforming the network architecture \cite{Fea14}. SDN technologies provide programmable data planes that can be configured from a remote controller platform. This control and data planes separation creates new opportunities to implement much more efficient traffic engineering policies than classical distributed protocols, since the (logically) centralized controller can take decisions on routing optimization exploiting a global view of the network and a flow level programmatic interface at data plane. Fault resilience mechanisms are among the most crucial traffic engineering instruments in operator networks since they insure quick reaction to connectivity failures with traffic rerouting.

So far, traffic engineering applications for SDN, and failure recovery solutions in particular, have received relatively little attention from the research community and networking industry which has focused mainly on other important areas related to security, load balancing, network slicing and service chaining. Not surprisingly, while SDN is becoming widely used in data centers where these applications are crucial, its adoption in operator networks is still rather limited. The support in current SDN implementations of features for failure recovery is currently rather weak and traditional technologies, like e.g. Multi-Protocol Label Switching (MPLS) Fast Reroute, are commonly considered for carrier networks more reliable.

The main reason for this gap in SDN solutions is that some traffic engineering applications, such as failure recovery, challenge the limits of the data plane abstraction that is the key element of any SDN architecture. OpenFlow is largely the most adopted abstraction for the data plane with its match-action rules in flow tables \cite{mckeown08}. Current OpenFlow abstraction presents some fundamental drawbacks that can prevent an efficient and performing implementation of traffic rerouting schemes. As a matter of fact, in OpenFlow adaptation and reconfiguration of forwarding rules (i.e. entries in the flow tables) in the data plane pipeline can only be performed by the remote controller, posing limitations on the granularity of the desired monitoring and traffic control due to the overhead and latency required. 

We believe that failure detection and reaction can be better handled locally in the switches assuming different sets of forwarding rules that can be applied according to the observed network state. This can be done retaining the logically centralized approach of SDN to programmability if we fully expose to application developers in the network controller both the state detection mechanism (i.e. link/node availabilities) and the sets of rules for the different states. The extension of the OpenFlow abstraction to stateful data planes has recently attracted the interest of the SDN research community: OpenState \cite{bianchi14} (proposed by some of the authors), FAST \cite{Mos14}, and the ``learn'' action of Open vSwitch \cite{ovs} are the main examples.

In this paper we propose SPIDER\footnote{\textbf{S}tateful \textbf{P}rogrammable fa\textbf{I}lure \textbf{DE}tection and \textbf{R}ecovery}, a fault resilient SDN pipeline design that allows the implementation of failure recovery policies with fully programmable detection and reaction mechanisms in the switches. SPIDER is based on stateful data planes and it provides guaranteed short failure detection and recovery delays, with a configurable trade off between overhead and failover responsiveness. We present a prototype implementation of SPIDER based on OpenState prototype switch and controller \cite{openstatehomepage}, and its performance evaluation on some example network topologies.

The paper is organized as follows. In Section \ref{sec:related_work} we discuss related work, while in Section \ref{sec:openstate} we review the characteristics of stateful data planes for SDN. In Section \ref{sec:approach} we introduce SPIDER approach and in Section \ref{sec:implementation} we outline its pipeline design and prototype implementation. Section \ref{sec:experimental_results} provides experimental results, while Section \ref{sec:conclusion} concludes the paper with our final remarks.

\section{Related Work}
\label{sec:related_work}

The concern of quickly recovering from failures in SDN has been already explored by the research community with the general goal of making SDN more reliable by reducing the need of switches to rely on the external controller to establish an alternative path. Sharma et al. in \cite{sharma13} shows how hard it is to obtain carrier grade recovery times ($<$50ms) when relying on a controller-based restoration approach in large OpenFlow networks. To overcome to such an issue, the authors propose also a proactive protection scheme based on a BFD daemon running in the switch and integrated with the OpenFlow Fast-failover group type, obtaining recovery times within 50ms. Similarly, Van Adrichem et al. shows in \cite{van14} how by carefully configuring the BFD process already compiled in Open vSwitch, it is possible to obtain recovery times of few ms. The case of protection switching is also explored by Kempf et al. in \cite{kempf12}, here the authors propose an end-to-end protection scheme based on an extended version of OpenFlow 1.1 to implement a specialized monitoring function to reduce processing load at the controller. Sgambelluri et al. proposed in \cite{sgambelluri13} a segment-protection approach based on pre-installed backup paths. Also in this case, OpenFlow is extended in order to enable switches to locally react to failures by auto-rejecting flow entries of the failed interface. The concern of reducing load at the controller is also addressed by Lee at al. in \cite{lee14}. A controller-based monitoring scheme and optimization model is proposed in order to reduce the number of monitoring iterations that the controller must perform to check all links. A completely different and more theoretical approach based on graph search algorithms is proposed by Borokhovich et al. in \cite{borokhovich14}. In this case the backup paths are not known in advance, but a solution based on the OpenFlow fast-failover scheme is proposed along an algorithm to randomly try new ports to reach traffic demands' destination.

Our work extends an earlier paper \cite{cascone15} were we first describe an OpenState-based behavioral model to perform fast reroute. In addition to that, we describe here for the first time a solution to provide programmable failure detection, including results on flow entries analysis, packet loss and heartbeat overhead. Finally, to the best of our knowledge, we are unaware of other prior work towards the use of programmable stateful data plane abstractions to implement both failure detection and recovery schemes.

\section{Stateful data planes}
\label{sec:openstate}

OpenFlow describes a stateless data plane abstraction for packet forwarding. Network states are maintained only at the controller, which in turn, based on a reactive approach, updates the flow table as a consequence of events such as the arrival of new flows, topology changes, or monitoring-based events triggered by the periodic polling of flow table statistics. We argue that improved scalability and responsiveness of network applications could be offered by adopting a stateful proactive abstraction, where switches are pre-provisioned with different sets of forwarding behaviors (i.e. flow entries), dynamically activated/deactived as a consequence of packet-level events and timers, and based on states maintained by the swich itself. OpenState \cite{bianchi14}, FAST \cite{Mos14} and OVS \cite{ovs} are example of such an abstraction supporting stateful forwarding. OpenState and FAST offers an explicit support to programming data plane state machines by defining dedicated structures such as state tables and primitives for state transition. OVS in turn, provides implicit support to stateful forwarding thanks to a special ``learn'' action (not currently supported in the OpenFlow specification) that allows the creation at run-time of new flow entries as a consequence of a packets matching existing ones.%\footnote{For a detailed description of OVS's stateful primitives, and an example on how to program stateful applications such a MAC learning switch, please refer to \cite{ovs-tutorial}.}.

We choose to base our design and implementation on OpenState for two reasons. First because, in our belief, OpenState offers a simple stateful forwarding abstraction that better serves the purpose of describing the behavioral model implemented by SPIDER in the form of Finite State Machines (FSMs). Indeed, while OVS's ``learn'' action could be used in principle to equivalently compile SPIDER features at data plane, it would require a less trivial effort in describing its design. Regarding FAST, although it provides a very similar abstraction to OpenState, unfortunately, as of today there is no publicly available implementation that we can use to implement and test SPIDER. Our second reason is that SPIDER is built on the assumption that updates of the forwarding state are possible at wire-speed, directly handled on the fast data path. The OpenState abstraction is also based on this assumption and its hardware experimental proof on a TCAM-based architecture was already addressed in \cite{pontarelli15}.

\begin{figure}
  \centering
  \includegraphics[width=1\columnwidth]{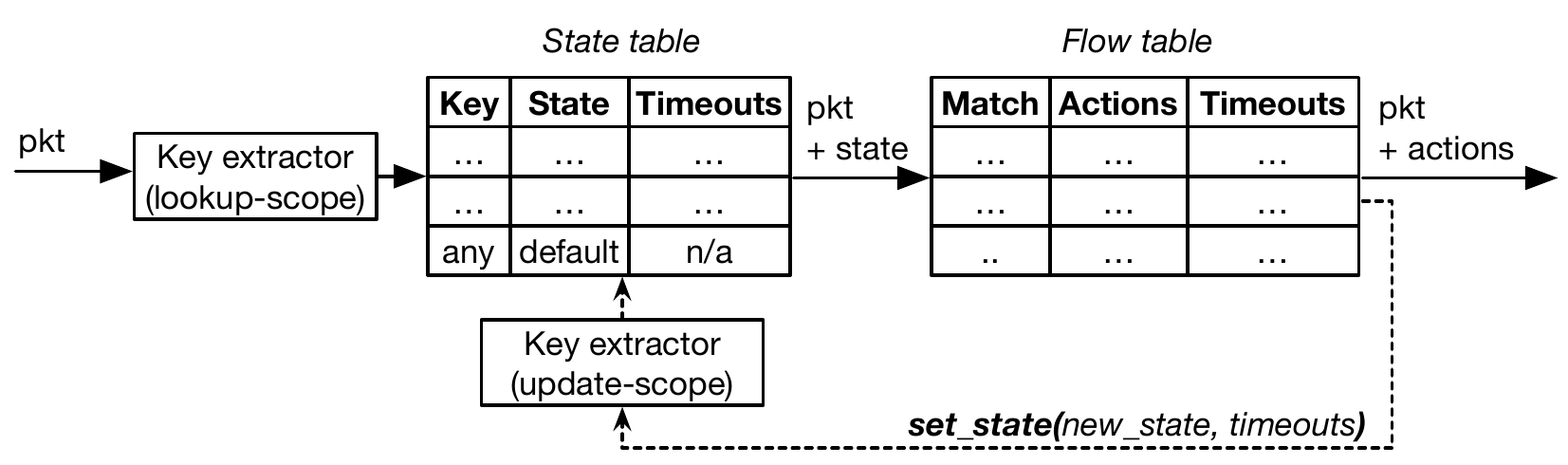}
  \caption{Architecture of a stage of the OpenState pipeline.}
  \label{fig:os-arch}
\vspace{-2mm}
\end{figure}

Before proceeding with the introduction of SPIDER, we consider it necessary to briefly summarize OpenState features, which are essential to define SPIDER in the rest of the paper\footnote{The features presented here are based on the most updated version of the OpenState v1.0 specification available at \cite{openstatehomepage}.}. Figure \ref{fig:os-arch} depicts the different elements of the OpenState pipeline. The legacy OpenFlow's flow table is preceded by a state table used to store ``flow states''. Each time a new packet is processed by the flow table, it is first matched against the state table. The matching is performed exactly (i.e. non-wildcard) on a flow key obtained using the fields defined by a ``lookup-scope'' (i.e. a list of OpenFlow's header fields identifier). The state table returns a ``default'' state if no state entry is matched by a given flow key, otherwise a different state is returned. The packet is then processed by the flow table, here flow entries can be defined to match on the state value returned by the state table. Moreover, a new ``set-state'' action is defined to insert/update entries in any state table of the pipeline. During a set-state action the state table is updated using a flow key optionally different from the one used in the lookup phase and defined by the ``update-scope'' (necessary when dealing with bidirectional flows). Finally, idle and hard state timeouts can be defined and are equivalent to those used in OpenFlow flow entries. A ``rollback state'' is associated to each timeout, and its value is used to update the state entry after the timeout expiration. Idle timeouts expires after a given entry is not matched by any packet for a given interval, while hard timeouts are expired counting from the instant the state entry has been inserted/updated. After configuring the lookup-scope, update-scope and the flow table, the state table is initially empty. It is then filled and updated based on the set-state actions defined in the flow table and executed as a consequence of packets matching flow entries in the flow table.

% By simply using the programming interface offered by OpenState, we are able to program the failure detection and recovery features offered by SPIDER without requiring any specialized packet processing or signaling primitives.

\begin{figure*}[]
  \centering
  \subfloat[][Normal (no failures)]{
    \includegraphics[width=0.31\textwidth]{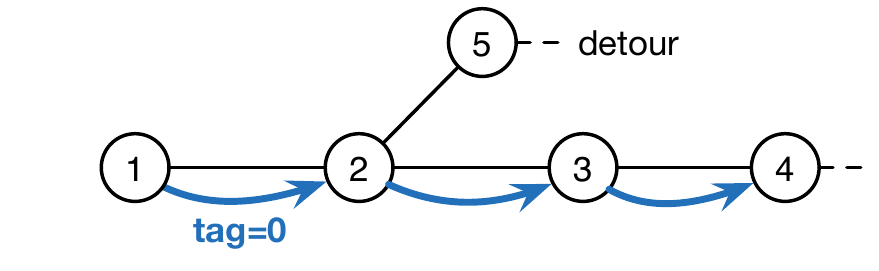}
    \label{fig:ex-normal}
  }
  \subfloat[][Local failover]{
    \includegraphics[width=0.3\textwidth]{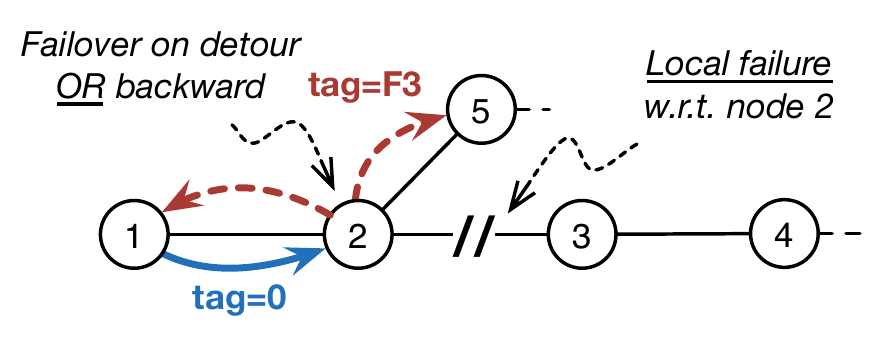}
    \label{fig:ex-local}
  }
  \subfloat[][Remote failover]{
    \includegraphics[width=0.3\textwidth]{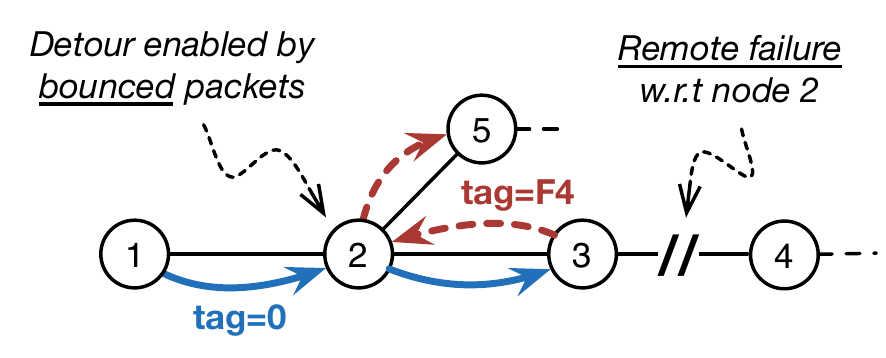}
    \label{fig:ex-remote}
  }
  \\
  \vspace{-5mm}
  \subfloat[][Heartbeat request/reply]{
    \includegraphics[width=0.3\textwidth]{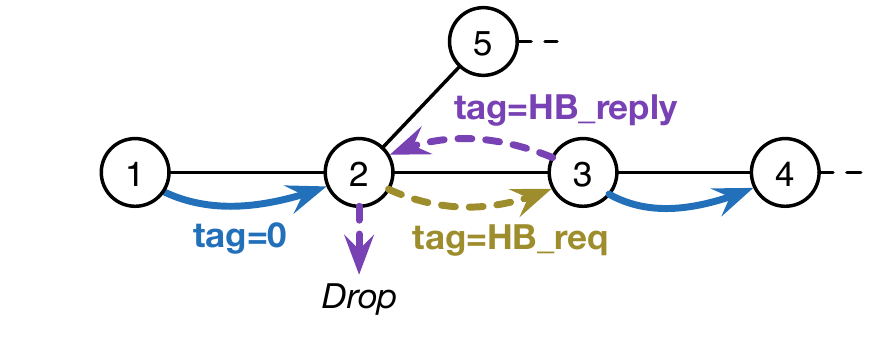}
    \label{fig:ex-heartbeat}
  }
  \subfloat[][Path probing]{
    \includegraphics[width=0.3\textwidth]{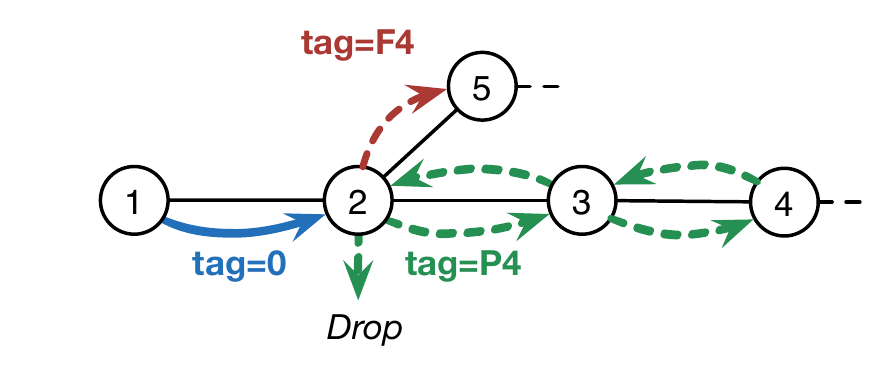}
    \label{fig:ex-probe}
  }
  \caption{Example of the different forwarding behaviors implemented by SPIDER.}
  \label{fig:ex-behaviors}
\vspace{-2mm}
\end{figure*}

\section{Approach sketch}
\label{sec:approach}

The features implemented by SPIDER are inspired by existing legacy protocols such as Bidirectional Forwarding Detection (BFD) \cite{rfc5880} and MPLS Fast Reroute \cite{rfc4090}. In this sense, SPIDER provides mechanisms to perform failure detection and instant rerouting of traffic demands without requiring the controller intervention. The controller interaction is needed only at boot time to provision switches' state tables and to fill flow tables with the different forwarding behaviors. No distributed protocols are required, instead the different forwarding behaviors are handled at data plane level by labeling packets with special tags and by using the stateful primitives introduced before.

\textbf{Backup path pre-planning.} SPIDER does not distinguish between node or link failures, instead we define with \s{Fi} a particular failure state of the network for which node $i$ is unreachable. Given another node $j$, we refer to the case of a ``local'' failure, when $j$ is directly connected (1 hop) to $i$, while we refer to a ``remote'' failure when node $i$ is not directly connected to $j$. In our design the controller must be provided with the topology of the network and a set of primary paths and backup paths for each demand. Backup paths must be provided for each possible \s{Fi} affecting the primary path of a given demand. A backup path for state \s{Fi} can share some of the primary path, but it is required to offer a detour (w.r.t primary path) around node $i$. In other words, even in the case of a link failure making $i$ unreachable from $j$, and even other links to $j$ might exist, we require that backup paths for \s{Fi} cannot use any of the links belonging to $i$. The reason of such a requirement is that, to guarantee very short ($<10ms$) failover delays, a characterization (link or node) of the failure is not possible without the active involvement of the controller or other type of slow signaling. For this reason SPIDER assumes always the worst case where node $i$ is down, hence it should be completely avoided. An example of problem formulation that can be used to compute an optimal set of such backup paths has been presented in \cite{Cap15}. Finally, if all backup paths are provided, SPIDER guarantees instantaneous protection from every single-failure \s{Fi} scenario, without requiring the controller to compute an alternative routing or to update flow tables. However, the unfortunate case of a second or multiple failures happening sequentially can be supported through the reactive intervention of the controller.

\textbf{Failure detection.} SPIDER uses tags (carried in a MPLS label) to distinguish between different forwarding behaviors and to perform failure detection and switch-to-switch failure signaling. Figure \ref{fig:ex-behaviors} depicts the different forwarding scenarios supported by SPIDER. When in normal conditions (i.e. no failures), packets entering the network are labeled with \s{tag=0} and routed through their primary path (Fig.~\ref{fig:ex-normal}). To detect failures, SPIDER doesn't rely on any switch-dependent feature such OpenFlow's Fast-failover, instead it provides a simple detection scheme based on the exchange of bidirectional ``heartbeat'' packets. We assume that as long as packet are received from a given port, that port can be also used to reliably transmit other packets. When no packets are received for a given interval, a node can request its neighbor to send an heartbeat. As shown in Fig.~\ref{fig:ex-heartbeat}, heartbeat can be requested by labeling any data packet with \s{tag=HB\_req}. A node receiving such a packet will perform 2 operations: i) set back \s{tag=0} and transmit the packet towards the next hop and ii) create a copy with \s{tag=HB\_reply} and send it back on the same input. In this way the node that requested the heartbeat will know that its neighbor is still reachable. Heartbeat are requested only when the received packet rate drops below a given threshold. If no packets (either data or heartbeat) are received for more than a given timeout, the port is declared \s{DOWN}. The state of the port will be set back to \s{UP} as soon as packets will be received again on that port.

\textbf{Fast reroute.} When a port is declared \s{DOWN}, meaning a local failure situation towards a neighbor node $i$, incoming packets are labeled with \s{tag=Fi} and sent to an alternative port (Fig.~\ref{fig:ex-local}), this could be a secondary port belonging to a detour or the same input port where the packet was received. In the last case we refer to a ``bounced'' packet. Bounced packets are used by SPIDER to signal a remote failure situation. Indeed, they are forwarded back along their primary path until they reach a node able to forward them along a detour. In Fig.~\ref{fig:ex-remote}, when node 2 receives a bounced packet with \s{tag=F4}, it updates the state of that demand to \s{F4} and forwards the packet along a detour. Given the stateful nature of SPIDER, state \s{F4} is maintained by node 2, meaning that all future packets of that demand with \s{tag=0}, will be labeled with \s{tag=F4} and transmitted directly on the detour. In the example, we refer to node 2 as the ``reroute'' node of a given demand in state \s{F4}, while the portion of the path comprised between the node that detected the failure and the reroute node is called the ``bounce path''.

\textbf{Path probing.} Failures are temporary, for this reason SPIDER provides also a probe mechanism to establish the original forwarding as soon as the failure is resolved. When in state \s{Fi} the reroute nodes periodically generates probe packets to check the reachability of node $i$. As for heartbeat packets, probe packets are not forged by switches or the controller, instead, they are generated simply duplicating and labeling the same data packets processed by a reroute node. In Fig.~\ref{fig:ex-probe}, node 2 duplicates a \s{tag=0} packet. One copy is sent on the detour with \s{tag=F4}, while the other is labeled with \s{tag=Pi} and sent on the original primary path. If node $i$ becomes reachable again, it will bounce the probe packet towards the reroute node. The reception of a probe packet \s{Pi} from a node with a demand in state \s{Fi} will cause a state transition that will re-enable the normal forwarding on the primary path.

\textbf{Flowlet-aware failover.} SPIDER also addresses the issue of packet reordering that might occur during the remote failover. Indeed, in the example of Fig.~\ref{fig:ex-remote}, while new \s{tag=0} packets arrive at the reroute node, one or more (older) packets may be traveling backward on the bounce path. Such a situation might cause packets to be delivered out-of-order at the receiver, with the consequence of unnecessary throughput degradation for transport layer protocols such as TCP. For this reason SPIDER implements the ``Flowlet-aware'' forwarding scheme first introduced in \cite{kandula07}. While SPIDER is already aware of the failure, the same forwarding decision is maintained for packets belonging to the same burst; in other words, packets are still forwarded (and bounced) on the primary path until a given idle timeout (i.e. interval between bursts) is expired. Such a timeout can be evaluated by the controller at boot time and should be set as the maximum RTT measured over the bounce path of a given reroute node for state \s{Fi}. Effectively waiting for such an amount of time before enabling the detour, maximizes the probability that no more packets are traveling back on the bounce path, thus minimizing the risk of mis-ordered packet at the receiver.

\section{Implementation}
\label{sec:implementation}

\begin{figure}[]
  \centering
  \includegraphics[width=1\columnwidth]{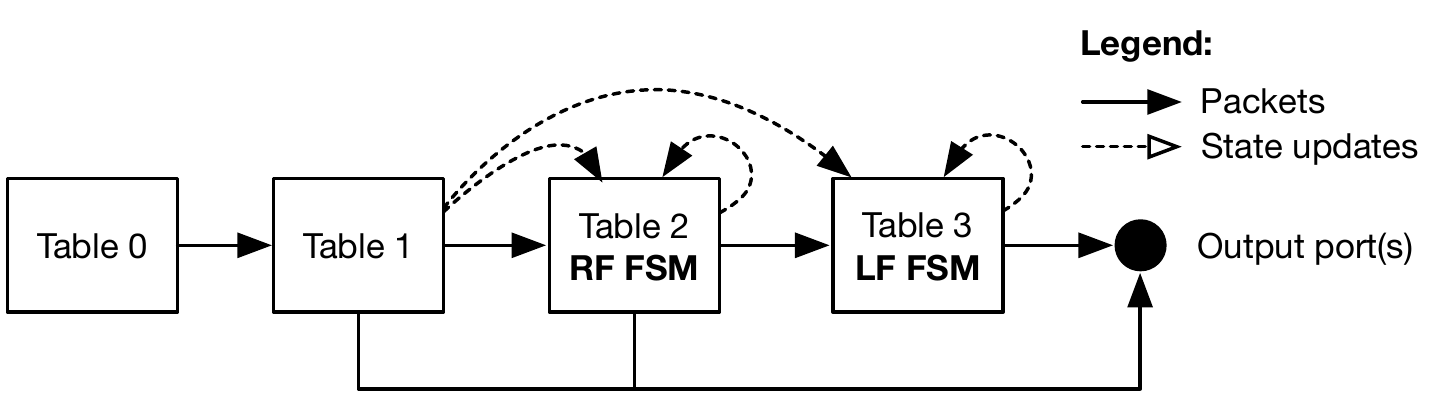}
  \caption{SPIDER pipeline architecture.}
  \label{fig:pipeline}
\vspace{-2mm}
\end{figure}

In this following section we present the architecture of the pipeline and the configuration of the flow tables necessary to implement SPIDER. The pipeline (Fig.~\ref{fig:pipeline}) is based on 4 different flow tables. An incoming packet is first processed by table 0 and 1. These two blocks perform only stateless forwarding (i.e. legacy OpenFlow), which features will be described later. The packet is then processed by stateful tables 2 and 3. These tables implements respectively the Remote Failover (RF) FSM, and the Local failover (LF) FSM described later. Packets are always processed by table 2 which is responsible for rerouting packets when the primary path of a given demand is affected by a remote failure. If no remote failure has been signaled to table 2, packets are submitted to table 3 which handles the failover in the case of local failures (i.e. directly seen by local ports). State updates in table 2 are triggered by bounced packets, while table 3 implements the heartbeat-based detection mechanisms introduced in Section \ref{sec:approach}. Although table 1 is stateless and for this reason doesn't need to maintain any state, it is responsible for triggering state updates on tables 2 and 3.

\textbf{Table 0.} It performs the following stateless processing before submitting packets to table 1:
\begin{itemize}
  \item For packets received from an edge port (i.e. directly connected to a host): push an initial MPLS label to store the tag.
  \item For packets received from a transport port (i.e. connected to another switch): write the input port in the metadata field (used later to trigger state updates from table 1).
  % \item All other packets are directly submitted to table 1.
\end{itemize}

\textbf{Table 1.} It handles the processing of those packets which requires only stateless forwarding, i.e. which forwarding behavior doesn't depend on states:
\begin{itemize}
  \item Data packets received at a edge port: set \s{tag=0}%\footnote{Because of a limitation of the current OpenFlow specification, the MPLS label cannot be pushed and set on the same table.}
  , then submit it to the next table.
  \item Data packets received at the last node of the primary path: pop the MPLS label, then directly transmitted on the corresponding output port (where the destination host is located).
  \item Packets with \s{tag=Fi}: directly transmitted on the detour port (unique for each demand and value of \s{Fi}); set \s{tag=0} on the last node of the detour before re-entering the primary path. An exception is made for the reroute node of demand in state \s{Fi}, in this case the routing decision for these packets is stored in table 2.
  \item Heartbeat requests (\s{tag=HB\_req}): packets are duplicated, one copy is set with \s{tag=HB\_reply} and transmitted through the input port, the other is set with \s{tag=0} and then submitted to the next table.
  \item Heartbeat replies (\s{tag=HB\_reply}): dropped (used only to update the state on table 3).
  \item Probe packets (\s{tag=Pi}): directly transmitted on the corresponding output port belonging to the probe path (i.e. the primary path, unique for each demand and value of \s{Pi}) (e.g. Fig.~\ref{fig:ex-probe}).
\end{itemize}
Finally, table 1 performs the following state updates on table 2 and 3:
\begin{itemize}
  \item For all packets: a state update is performed on table 3 so to declare the port on which the packet has been received as \s{UP}.
  \item Only for probe packets: a state update is performed on table 2 to transition a flow state from \s{Fi} to \s{Normal}.
\end{itemize}

\begin{figure}
  \centering
    \includegraphics[width=0.8\columnwidth]{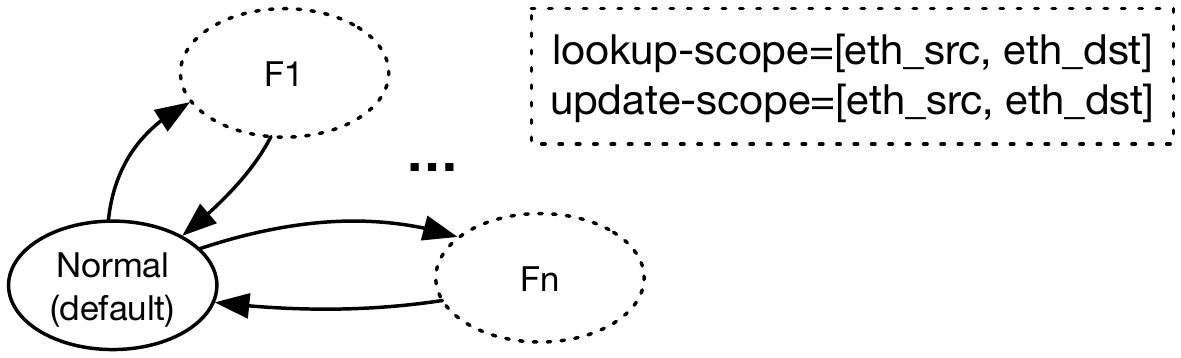}
    \caption{Macro states of the Remote Failover FSM}
    \label{fig:rf-fsm-simple}
\vspace{-2mm}
\end{figure}

\begin{figure}
  \centering
    \includegraphics[width=1\columnwidth]{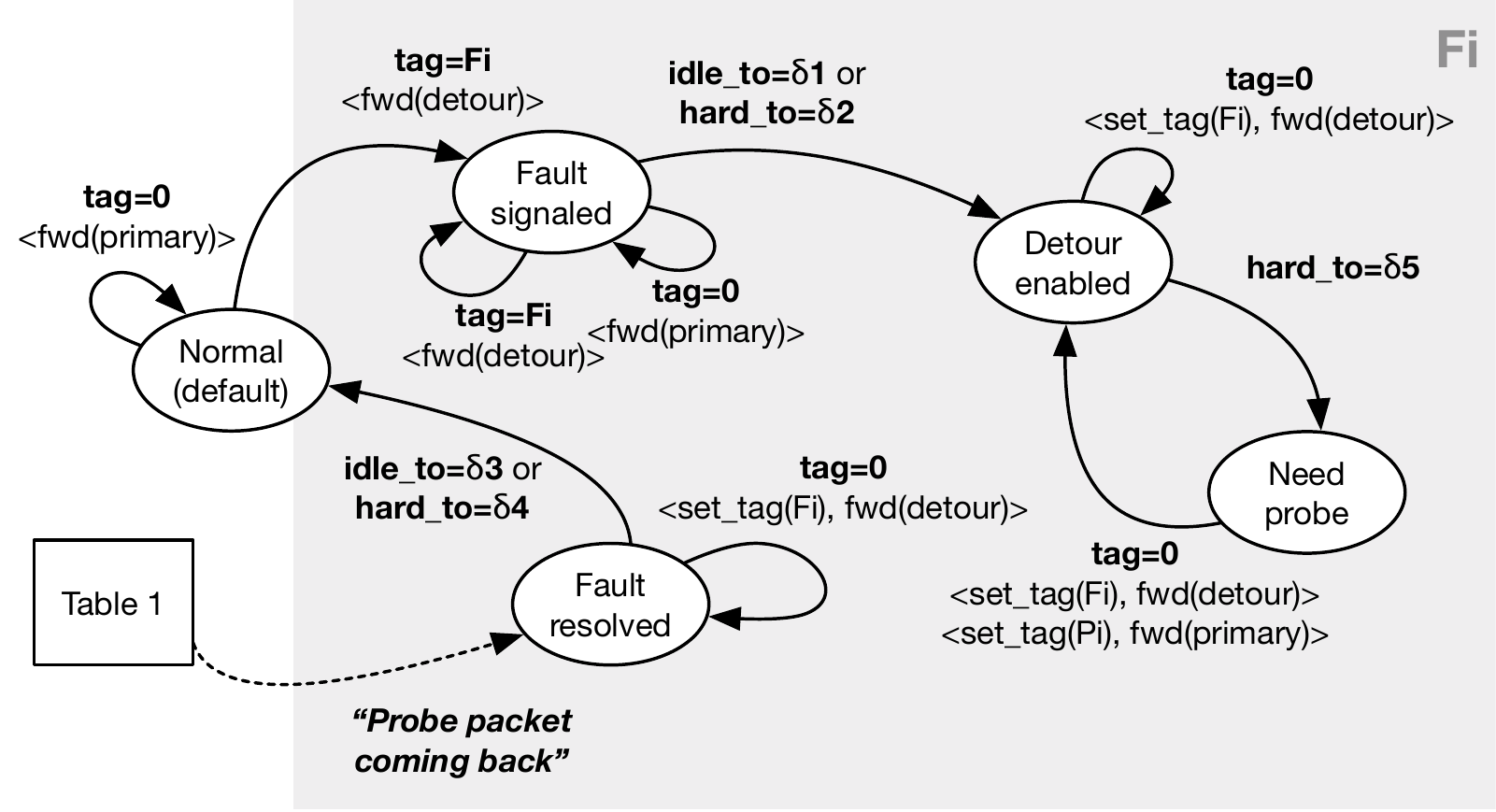}
    \caption{Detail of the macro state \s{Fi} for the Remote Failover FSM.}
    \label{fig:rf-fsm-detailed}
\vspace{-2mm}
\end{figure}

\textbf{Table 2 (Remote Failover FSM).} Figure~\ref{fig:rf-fsm-simple} shows a simplified version of the FSM. A state is maintained for each different traffic demand served by the switch. As outlined by the lookup and update scopes, in this case the origin-destination demands are identified by the tuple of Ethernet source and destination address, a programmer might specify different aggregation fields to describe the demands (e.g. IP source/destination tuple, or the 4-tuple transport layer protocol). Given the support for only single-failure scenarios, transitions between macro states \s{Fi} are not allowed (state must be set to \s{Normal} before transitioning to another state \s{Fi}).
Figure \ref{fig:rf-fsm-detailed} depicts a detailed version of the Remote Failover FSM with macro state \s{Fi} exploded. At boot time the state of each demand is set to the default value \s{Normal}. Upon reception of a bounced packet with \s{tag=Fi}, the latter is forwarded on the detour and state set to \s{Fault signaled}. The flowlet-aware routing scheme presented before, is here implemented by means of state timeouts. When in state \s{Fault signaled}, packets arriving with \s{tag=0} (i.e. from the source node) are still forwarded on the primary path.%, bounced with \s{tag=Fi} at the detect node, and finally forwarded on the detour.
This behavior is maintained until the expiration of the idle timeout $\delta_1$, i.e. after no packets of that demand have been received for a $\delta_1$ interval, which should be set equal to the RTT measured on the bounce path\footnote{Such a feature requires the support for very short timeouts. OpenState v1.0 currently define state timeouts with microseconds resolution.}.To avoid a situation where the demand remains locked in state \s{Fault signaled}, an hard timeout $\delta_2>\delta_1$ is set so that the next state \s{Detour ready} is always reached after at most a $\delta_2$ interval.
When in state \s{Detour enabled}, packets are set with \s{tag=Fi} and transmitted directly on the detour. In this state an hard timeout $\delta_5$ assures the periodic transmission of probe packets on the primary path. The first packet matched when in state \s{Need probe} is duplicated: one copy is sent on the detour towards its destination, another copy is set with \s{tag=Pi} and sent to node $i$ through the original primary path of the demand. If node $i$ becomes reachable again, it responds to the probe by bouncing the packet (\s{tag=Pi} is maintained) to the reroute node that originated it. The match of the probe packet at table 1 of the reroute node will trigger a %\footnote{We refer to the term forced as these updates cannot be described as classic state transitions originating from a given state. In this case the state is set regardless of the current state of a given instance of the FSM of table 2. We prefer to break the formalism and simplify the description of the behavioral model for the specific case of a multi-table stateful pipeline.}
reset of the Remote Failure FSM to state \s{Fault resolved}. %Otherwise, if node $i$ is still unreachable, meaning that the failure has not been resolved, probe packets will be lost along their way.
When in state \s{Fault resolved}, the same flowlet-aware routing scheme of state \s{Fault signaled} is applied. In this case an idle and hard timeout are set in order to maintain the current routing (on the detour) until the end of the transmission of the current burst of packets. In this case $\delta_3$ must be set to the maximum delay difference between the primary and backup path. After the expiration of $\delta_3$ or $\delta_4$, the state is set back to \s{Normal}, hence the transmission on the detour stops and packets are submitted to table 3 to be forwarded on their primary port.

\begin{figure}
  \centering
    \includegraphics[width=1\columnwidth]{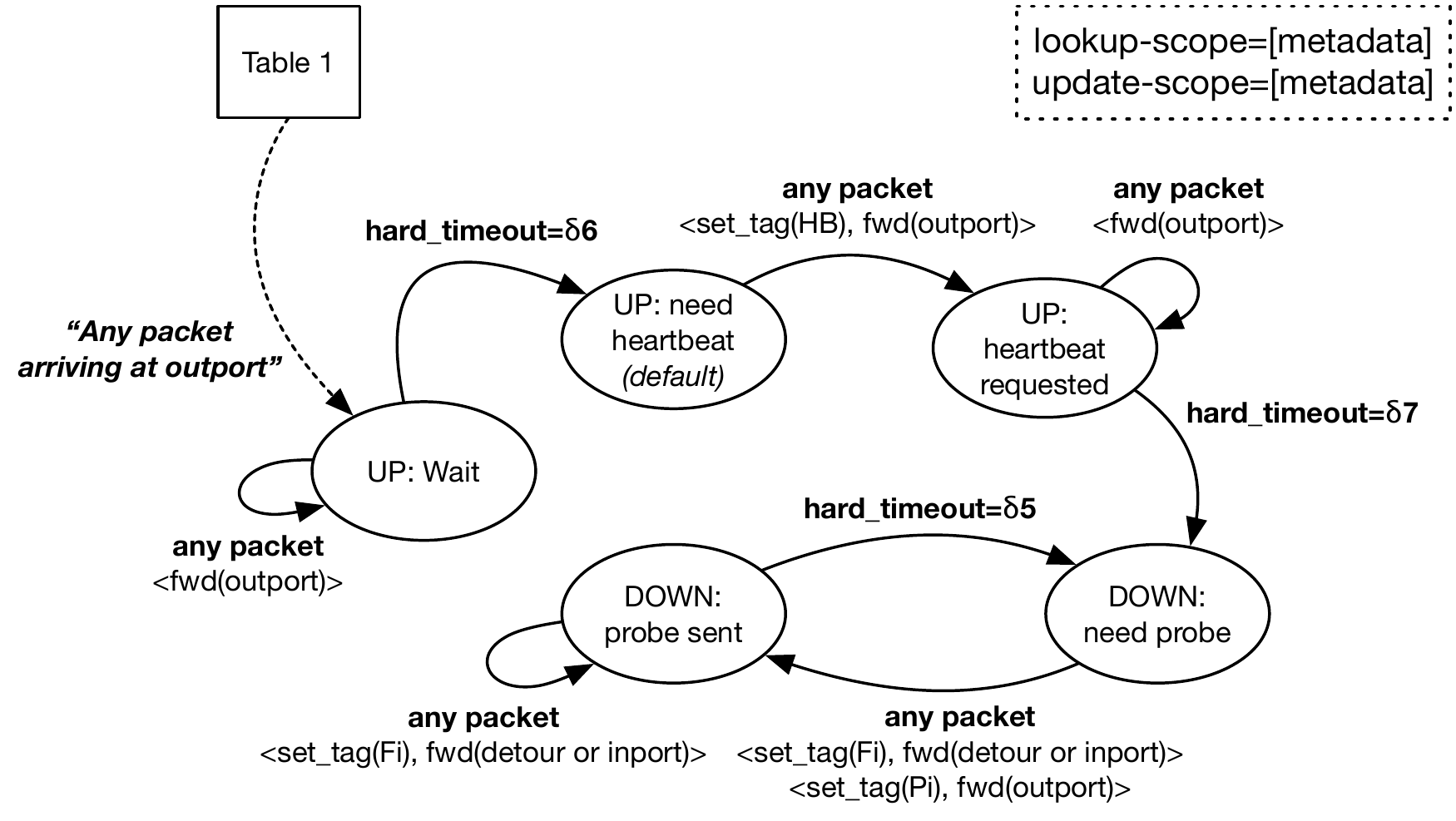}
    \caption{Mealy machine implemented by the LF table}
    \label{fig:lf-fsm}
\vspace{-2mm}
\end{figure}

\textbf{Table 3 (Local Failover FSM).} Figure \ref{fig:lf-fsm} depicts the FSM implemented by this table. Here flows are aggregated per output port (encoded in the metadata field)%\footnote{In our current implementation based on OpenFlow 1.3 matching on the outport is not supported, for this reason we use the metadata field to carry this information across tables.}
, meaning that all packets destined to the same port will share the state. This FSM has two macro states, namely \s{UP} and \s{DOWN}. When in state \s{DOWN}, packets are forwarded to an alternative port (detour or input port in case of bounced packets, according to the pre-planned backup strategy). %While the primary output port is set by table 2, the backup port is specific for each demand and stored in table 3.
At boot time all flows are optimistically set in default state \s{UP: need heartbeat}, meaning that an heartbeat packet must be generated and a reply received, so that the port keeps being declared UP. Indeed, the first packet matched in this state will be sent with \s{tag=HB\_req} and the state updated to \s{UP: heartbeat requested}. While in this state, packets will be transmitted on the primary output port, until an hard timeout $\delta_7$ expires, in which case the port will be declared \s{DOWN}. The timeout $\delta_7$ represents the maximum interval admitted between the generation of the heartbeat request and the reception of the corresponding reply. Every time a packet (either a data, probe or heartbeat) is received at table 1 the state of that port is reset to \s{UP: wait}. The Local Failover FSM will stay in this state for an interval $\delta_6$ (hard timeout), after which the state will be set back to \s{UP: need heartbeat}. Hence, $\delta_6$ represents the inverse of the minimum received packet rate required for a given port to avoid the generation of heartbeats. If the timeout $\delta_7$ expires, the port is declared \s{DOWN}. Here, packets will be tagged with \s{Fi} (where $i$ represents the failed node directly connected to the port) and forwarded on an alternative port. Similarly to the Remote Failover FSM, an hard timeout $\delta_5$ assures that probe packets will be generated even when the port is declared \s{DOWN}.% By doing so, when the failure will be resolved, the reception of the probe packet \s{Pi} by table 1 will force a state transition to \s{UP: wait}, in which case the transmission will be re-established on the primary output port.

In conclusion, Table \ref{table:timeouts} summarizes the different timeouts used in the SPIDER pipeline. We emphasize how, by tweaking these values, a programmer can explicitly control and impose i) a precise detection delay for a given port ($\delta_6$ + $\delta_7$), ii) the level of traffic overhead caused by probe packets of a given demand ($\delta_5$ and $\delta_6$), the risk of packets reordering in the case of a remote failover ($\delta_1$, $\delta_2$, $\delta_3$, and $\delta_4$). Experimental results based on these parameters are presented in the following section.

\begin{table*}[]
\centering
\caption{Summary of the configurable timeouts of the SPIDER pipeline}
\label{table:timeouts}
\begin{tabular}{@{}p{0.05\textwidth}p{0.03\textwidth}p{0.355\textwidth}p{0.475\textwidth}@{}}
\toprule
\textbf{Timeout} & \textbf{Type} & \textbf{Description} & \textbf{Value} \\ \midrule
$\delta_1$ & Idle & Flowlet idle timeout before switching packets from the primary path to the detour & Maximum RTT measured on the bounce path for a specific demand and \s{Fi} \\ \hline
$\delta_2$ & Hard & Maximum interval admitted for the previous case before enabling the detour & $>\delta_1$ \\ \hline
$\delta_3$ & Idle & Flowlet idle timeout before switching packets from the detour to the primary path & Maximum end-to-end delay difference between the backup path and the primary path \\ \hline
$\delta_4$ & Hard & Maximum interval admitted for the previous case before re-enabling the primary path & $>\delta_3$ \\ \hline
$\delta_5$ & Hard & Probe generation timeout & Arbitrary interval between each periodic check of the primary path in case of remote failure \\ \hline
$\delta_6$ & Hard & Heartbeat requests generation timeout & Inverse of the minimum rx rate for a given port before the generation of heartbeat requests and the corresponding replies \\ \hline
$\delta_7$ & Hard & Heartbeat reply timeout before the port is declared down & Maximum RTT for heartbeat requests/replies between two specific nodes (1 hop) \\ \bottomrule
\end{tabular}
\end{table*}

\subsection*{Experimental implementation}

SPIDER has been implemented using a modified version of the OpenFlow Ryu controller \cite{ryu} extended in order to support OpenState \cite{openstatehomepage}. SPIDER source code is available at \cite{SPIDER_repository}. For the experimental performance evaluation we used an emulated testbed based on Mininet \cite{mininet} using a version of the CPqD OpenFlow 1.3 softswitch \cite{ofsoftswitch13} as well extended with OpenState support \cite{openstatehomepage}.

\begin{table}[]
\centering
\caption{Number of flow entries per node.}
\label{table:BigO}
\begin{tabular}{llllllllllll}
\toprule
\textbf{Net} & \textbf{D} & \textbf{E} & \textbf{C} & \textbf{min} & \textbf{avg} & \textbf{max} & $E^2\times N$ \\ \midrule
5x5 & 240 & 16 & 9 & 402 & 727 & 934 & 6400 \\ \hline
6x6 & 380 & 20 & 16 & 497 & 1046 & 1490 & 14400 \\ \hline
7x7 & 552 & 24 & 25 & 720 & 1578 & 2280 & 28224 \\ \hline
8x8 & 756 & 28 & 36 & 998 & 2117 & 3523 & 50176 \\ \hline
9x9 & 992 & 32 & 49 & 1273 & 2744 & 4318 & 82944 \\ \hline
10x10 & 1260 & 36 & 64 & 1121 & 3421 & 5708 & 129600 \\ \hline
11x11 & 1560 & 40 & 81 & 1359 & 4061 & 7213 & 193600 \\ \hline
12x12 & 1892 & 44 & 100 & 1127 & 4915 & 9106 & 278784 \\ \hline
13x13 & 2256 & 48 & 121 & 1989 & 5977 & 10486 & 389376 \\ \hline
14x14 & 2652 & 52 & 144 & 1404 & 6892 & 14536 & 529984 \\ \hline
15x15 & 3080 & 56 & 169 & 3576 & 8171 & 15522 & 705600 \\ 

\bottomrule
\end{tabular}
\end{table}

\section{Performance Evaluation}
\label{sec:experimental_results}

\subsection{Flow entries analysis}

We start the analysis of SPIDER evaluating the resources required by a switch to implement the presented pipeline architecture in terms of flow table entries and memory required for states. We start by defining as $D$ the maximum number of demands served by a switch, $F$ the maximum number of failures that can affect a demand (i.e. length of the longest primary path), and $P$ the maximum number of ports of a switch. We can easily model the number of flow entries required by means of Big-O notation as $O(D \times F)$. Indeed, for table 0 the number of entries is equal to $P$; for table 1 in the worst case we have one entry per demand per fault ($D\times F$); for table 2 we always have exactly $7\times D\times F$, and for table 3 exactly $P\times (3+2\times D)$. In total, we have a number of entries order of $P+D\times F+D\times F+D\times P$ and then of $D\times F+D\times P$. Assuming $F>>P$ we can conclude that the number of entries is $O(D\times F)$.

If we want to evaluate the complexity according to network size, we can observe that in the worst case $F=N=E+C$, where $N$ is the number of nodes, $E$ the number of edge nodes and $C$ the number of core nodes. Assuming a path protection scheme, which is the most demanding in terms of rules since all the \s{Fi} states are managed by the ingress edge nodes, and a full traffic matrix, we have $D=E(E-1)\approx E^2$. In the worst case we have a single node managing all faults of all demands, where the primary path of each demand is the longest possible, thus $F=N$. In this case the number of entries will be $O(E^2\times N)$.

In Table \ref{table:BigO} we report the values for grid networks $n\times n$ where edge nodes are the outer ones of the grid and there is a traffic demand for each pair of edge nodes. In addition to the $O(E^2\times N)$ values, we include in the table the values per node (min, max, average) calculated for the case of end to end protection where the primary path is the shortest one (number of hops) and the backup path is the shortest node disjoint from the primary. The number of rules is generated according to the SPIDER implementation described in Section V and available at \cite{SPIDER_repository}. We can observe that even the max value is always much smaller than the values estimated by the complexity analysis. Obviously, for more efficient protection schemes based on a distributed handling of states $Fi$ (e.g. segment protection), we expect an even lower number of rules per node.

As far as the state table is concerned, table 2 for node $n$ needs $D_n$ entries, where $D_n$ is the number of demands for which $n$ is a reroute node. For the width of the table we need to consider the total number of possible states that is $1+4F_n$, where $F_n$ is the number of remote failures managed by $n$. Similarly, for stage 3 we have only $5$ possible states and a number of entries equal to $P$.

\subsection{Detection mechanism}

\begin{figure}[]
  \centering
    \scalebox{.55}{\input{fig/plots/packet_loss_probe_interval}}
    \caption{Packet loss (data rate $1000$ $pkt/sec$)}
    \label{fig:losses}
\vspace{-2mm}
\end{figure}
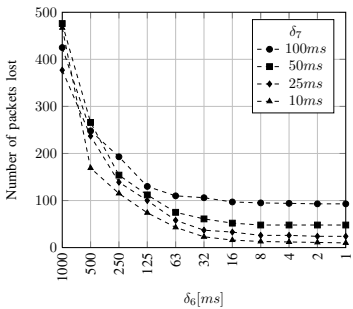

To evaluate the effectiveness of the SPIDER heartbeat-based detection mechanism, we have considered a simple experimental scenario of two nodes and a link with traffic of $1000$ $pkt/sec$ sent in one direction only. In Fig.~\ref{fig:losses} we show the number of packets lost after a link failure versus $\delta_6$ (heartbeat interval) and $\delta_7$ (heartbeat timeout). As expected, the number of losses decreases as the heartbeat interval and timeout decreases. In general, the number of dropped packets depends on the precise instant the failure occurs w.r.t. $\delta_6$ and $\delta_7$. The curves reported are obtained averaging the results of 10 different tries with failures reproduced at random instants.

\subsection{Overhead}

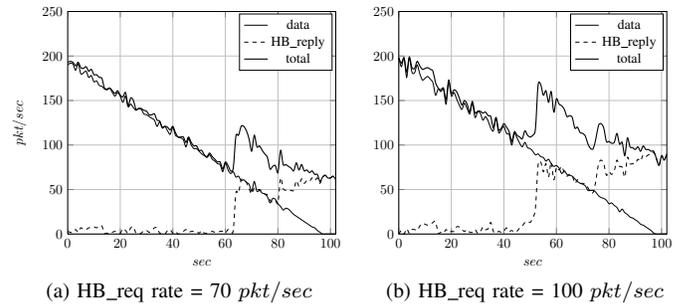
\begin{figure}[]
\centering
      % \subfloat[][HB\_req rate = 10 $ pkt/sec $]{
      %   \scalebox{.53}{\input{fig/plots/HB_OH_10}}
      %   \label{fig:p-oh-10}
      % }
      % %\hspace{2.7cm}
      % \subfloat[][HB\_req rate = 40 $ pkt/sec $]{
      %   \scalebox{.53}{\input{fig/plots/HB_OH_40}}
      %   \label{fig:p-oh-40}
      % }
      \subfloat[][HB\_req rate = 70 $ pkt/sec $]{
        \scalebox{.52}{\input{fig/plots/HB_OH_70}}
        \label{fig:p-oh-70}
      }
      %\hspace{2.7cm}
      \subfloat[][HB\_req rate = 100 $ pkt/sec $]{
        \scalebox{.52}{\input{fig/plots/HB_OH_100}}
        \label{fig:p-oh-100}
      }
\caption{Heartbeat overhead with decreasing data traffic 200-0 $ pkt/sec $ and heartbeat request rates (inverse of $\delta_6$) of 70, and 100 $ pkt/sec $.} \label{fig:p-oh}
\vspace{-2mm}
\end{figure}

Obviously, the price to pay for a small number of losses is the overhead due to heartbeat packets. However, SPIDER exploits the traffic in the reverse direction for failure detection, and this reduces the amount of heartbeat packets. For the same two nodes scenario in the previous section, we have evaluated the overhead caused when generating a decreasing traffic profile of 200 to 0 $ pkt/sec $, with different values of $\delta_6$. Results are reported in Fig.~\ref{fig:p-oh}.

We can see that, as long as the reverse traffic rate is higher than the heartbeat request rate ($1/\delta_6$), zero or low signaling overhead is observed. When the traffic rate decreases, the overhead due to heartbeats tends to compensate for the missing feedback packets up to the threshold. However, this overhead does not really affect the network performance since it is generated only when reverse traffic is low.

\subsection{Comparison with OpenFlow}

\begin{figure}[]
  \centering
  \subfloat[][]{
    \includegraphics[width=0.45\columnwidth]{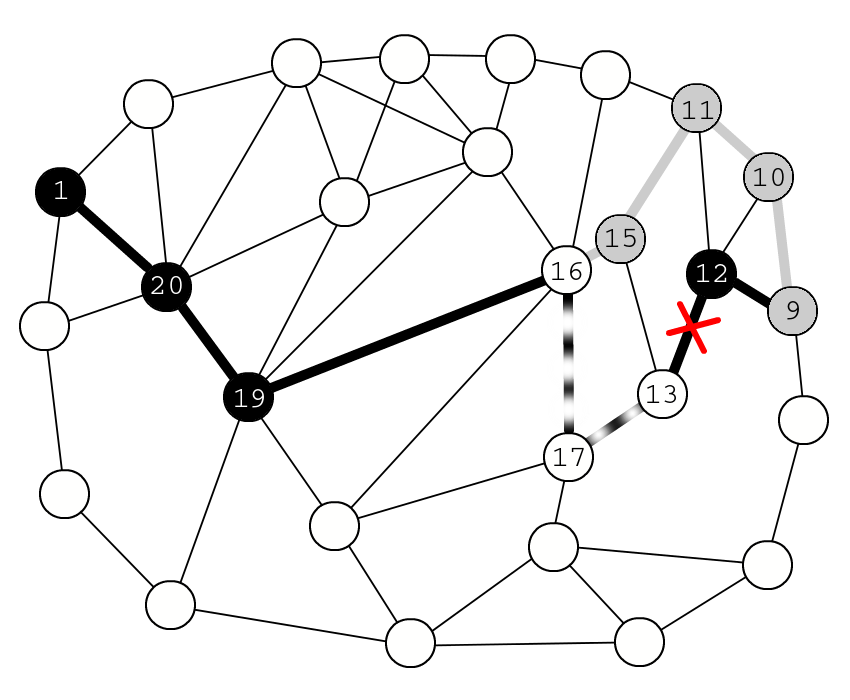}
    \label{fig:norway_topo}
  }
  \subfloat[][]{
    \scalebox{.51}{\input{fig/plots/packet_loss_FRIP_vs_OF}}
    \label{fig:SPIDERvsOF-chart}
  }
  \caption{Comparison with OpenFlow: (a) test topology used in experiments and (b) packet loss}
  \label{fig:SPIDERvsOF}
\vspace{-2mm}
\end{figure}

We now compare a SPIDER based solution with a strawman implementation corresponding to a reactive OpenFlow (OF) application able to modify the flow entries only when the failure is detected and notified to the controller. We have considered the network shown in Fig.~\ref{fig:norway_topo}. For the primary and backup paths, as well as the link failure indicated in the figure, we have considered an increasing number of demands with a fixed packet rate of $100$ $pkt/sec$ each one. For the OF case, we used the detection mechanism of the Fast-failover (FF) group type implemented by the CPqD softswitch, and different RTTs between the switch that detects the failure and the controller. For SPIDER we used a heartbeat interval ($\delta_6$) of 2 $ms$ and timeout ($\delta_7$) of 1 ms. For all the considered flows, no local backup path is available: in the SPIDER case the network is able to autonomously recover from the failure by bouncing packets on the primary path, while in the OF case the controller intervention is needed to restore connectivity.

The results obtained are shown in Fig.~\ref{fig:SPIDERvsOF-chart}. We can see that the losses in the case of SPIDER are always lower than OF. Note that, even if the heartbeat interval used is small, this is not actually an issue for the network since in the presence of reverse traffic the overhead is proportionally reduced so that it never affects the link available capacity. The value of the timeout actually depends on the maximum delay for heartbeat replies to be delivered, which in networks with high speed links mainly depends on propagation and can be set to low values by assigning maximum priority to heartbeat replies. In the case of OF, the number of losses increases as the RTT between switch and controller increases. Obviously, losses also increase with the number of demands since the total number of packets received before the controller installs the new rules increases as well.

\section{Conclusion}
\label{sec:conclusion}

In this paper we have presented SPIDER, a new approach to failure recovery in SDN that provides a fully programmable abstraction to application developer for the definition of the re-routing policies and for the management of the failure detection mechanism. The use of recently proposed stateful data planes, allows to execute the programmed failure reaction behaviors directly in the switches, minimizing the recovery delay and guaranteeing the failover even when the controller is not reachable. We believe that the proposed approach can close one of the gaps between the required and supported features that at the moment are slowing down the adoption of SDN in carrier grade networks for telco operators.

SPIDER has been implemented using OpenState. The prototype implementation (code is made available at \cite{SPIDER_repository}) has been used to validate the proposed scheme and to experimentally assess its basic performance in a few example scenarios. The results have shown the potential advantages of SPIDER with respect to fully centralized applications where the controller is notified of failure events and is required to modify all affected forwarding rules.

\section*{Acknowledgment}
This work has been partly funded by the EU in the context of the ``BEBA'' project \cite{beba} (Grant Agreement: 644122).

\newpage
\bibliographystyle{IEEEtran}
\bibliography{99-biblio}

% that's all folks
\end{document}

%% file: fig/plots/packet_loss_probe_interval.tex
\begin{tikzpicture}
\begin{axis}[
    grid = major,
    xmin = 1, xmax = 11,
    ymin = 0, ymax = 500,
    ylabel={Number of packets lost},
    ylabel style={yshift=-0.1em},
    legend style ={at={(0.67,0.97)},anchor=north west, draw=black,fill=white,align=left},
    xlabel={$ \delta_6 [ ms ] $},
    xlabel style={yshift=-1.2em},
    xtick={1,2,3,4,5,6,7,8,9,10,11},
    xticklabels={1000 , 500 , 250 , 125 , 63 , 32 , 16 , 8 , 4 , 2 , 1},
    x tick label style={rotate=90,anchor=east}
]
\addlegendimage{empty legend}
\addlegendentry{\hspace{-.6cm}\vspace{-.6cm} $ \delta_7 $}
\addplot[color=black,dashed,mark=*,mark options={solid}]
coordinates{
    (1,425)
    (2,248) 
    (3,193)
    (4,130)
    (5,110)
    (6,106)
    (7,97)
    (8,95)
    (9,94)
    (10,93)
    (11,93)
    };
\addlegendentry{$100 ms$}
\addplot[color=black,dashed,mark=square*,mark options={solid}]
coordinates{
    (1,476)
    (2,266) 
    (3,154)
    (4,112)
    (5,75)
    (6,61)
    (7,52)
    (8,48)
    (9,48)
    (10,48)
    (11,48)
};
\addlegendentry{$50 ms$}
\addplot[color=black,dashed,mark=diamond*,mark options={solid}]
coordinates{
    (1,377)
    (2,237) 
    (3,139)
    (4,100)
    (5,58)
    (6,37)
    (7,33)
    (8,26)
    (9,26)
    (10,24)
    (11,24)
};
\addlegendentry{$25 ms$}
\addplot[color=black,dashed,mark=triangle*,mark options={solid}]
coordinates{
    (1,467)
    (2,169) 
    (3,115)
    (4,74)
    (5,43)
    (6,23)
    (7,16)
    (8,13)
    (9,12)
    (10,11)
    (11,10)
};
\addlegendentry{$10 ms$}
\end{axis}
\end{tikzpicture}

%% file: fig/plots/HB_OH_70.tex
\begin{tikzpicture}
\begin{axis}[
    grid=major,
    xmin = 0, xmax = 102,
    ymin = 0, ymax = 250,
    ylabel={$ pkt/sec $},
    legend style ={ at={(0.64,0.99)},anchor=north west, draw=black,fill=white,align=left},
    xlabel={$ sec $}
]
\addplot[
   color=black,
   solid,
   thin,
   smooth
   ]
coordinates{
 (0, 190)
 (1, 192)
 (2, 191)
 (3, 184)
 (4, 191)
 (5, 180)
 (6, 181)
 (7, 183)
 (8, 176)
 (9, 178)
 (10, 171)
 (11, 172)
 (12, 167)
 (13, 166)
 (14, 163)
 (15, 162)
 (16, 164)
 (17, 159)
 (18, 155)
 (19, 159)
 (20, 155)
 (21, 148)
 (22, 155)
 (23, 143)
 (24, 146)
 (25, 142)
 (26, 139)
 (27, 141)
 (28, 140)
 (29, 135)
 (30, 134)
 (31, 132)
 (32, 130)
 (33, 128)
 (34, 121)
 (35, 126)
 (36, 119)
 (37, 121)
 (38, 116)
 (39, 112)
 (40, 120)
 (41, 111)
 (42, 109)
 (43, 108)
 (44, 106)
 (45, 98)
 (46, 107)
 (47, 100)
 (48, 96)
 (49, 96)
 (50, 93)
 (51, 90)
 (52, 88)
 (53, 86)
 (54, 89)
 (55, 84)
 (56, 82)
 (57, 79)
 (58, 78)
 (59, 70)
 (60, 80)
 (61, 72)
 (62, 68)
 (63, 70)
 (64, 64)
 (65, 65)
 (66, 60)
 (67, 61)
 (68, 58)
 (69, 56)
 (70, 48)
 (71, 59)
 (72, 48)
 (73, 49)
 (74, 46)
 (75, 45)
 (76, 42)
 (77, 35)
 (78, 40)
 (79, 39)
 (80, 34)
 (81, 31)
 (82, 27)
 (83, 29)
 (84, 28)
 (85, 25)
 (86, 22)
 (87, 20)
 (88, 18)
 (89, 16)
 (90, 14)
 (91, 12)
 (92, 10)
 (93, 8)
 (94, 7)
 (95, 4)
 (96, 2)
 (97, 0)
 (98, 0)
 (99, 0)
 (100, 0)
 (101, 0)
 (102, 0)
};
\addlegendentry{data}
\addplot[
   color=black,
   dash pattern=on 3pt off 3pt on 3pt off 3pt,
   smooth
   ]
coordinates{
 (0, 3)
 (1, 2)
 (2, 2)
 (3, 5)
 (4, 2)
 (5, 7)
 (6, 5)
 (7, 5)
 (8, 7)
 (9, 7)
 (10, 8)
 (11, 7)
 (12, 9)
 (13, 9)
 (14, 1)
 (15, 1)
 (16, 0)
 (17, 2)
 (18, 3)
 (19, 2)
 (20, 1)
 (21, 6)
 (22, 4)
 (23, 1)
 (24, 7)
 (25, 8)
 (26, 1)
 (27, 0)
 (28, 1)
 (29, 4)
 (30, 4)
 (31, 2)
 (32, 4)
 (33, 7)
 (34, 7)
 (35, 1)
 (36, 1)
 (37, 0)
 (38, 4)
 (39, 5)
 (40, 1)
 (41, 3)
 (42, 0)
 (43, 1)
 (44, 3)
 (45, 5)
 (46, 2)
 (47, 3)
 (48, 3)
 (49, 3)
 (50, 1)
 (51, 4)
 (52, 1)
 (53, 1)
 (54, 2)
 (55, 3)
 (56, 2)
 (57, 2)
 (58, 3)
 (59, 3)
 (60, 1)
 (61, 5)
 (62, 3)
 (63, 4)
 (64, 46)
 (65, 46)
 (66, 61)
 (67, 60)
 (68, 58)
 (69, 51)
 (70, 45)
 (71, 52)
 (72, 50)
 (73, 50)
 (74, 48)
 (75, 45)
 (76, 44)
 (77, 38)
 (78, 36)
 (79, 38)
 (80, 35)
 (81, 63)
 (82, 51)
 (83, 49)
 (84, 48)
 (85, 49)
 (86, 49)
 (87, 60)
 (88, 54)
 (89, 62)
 (90, 57)
 (91, 60)
 (92, 59)
 (93, 63)
 (94, 60)
 (95, 63)
 (96, 57)
 (97, 64)
 (98, 66)
 (99, 63)
 (100, 63)
 (101, 65)
 (102, 61)
};
\addlegendentry{HB\_reply}
\addplot[
   color=black,
   solid,
   smooth,
   thick
   ]
coordinates{
 (0, 193)
 (1, 194)
 (2, 193)
 (3, 189)
 (4, 193)
 (5, 187)
 (6, 186)
 (7, 188)
 (8, 183)
 (9, 185)
 (10, 179)
 (11, 179)
 (12, 176)
 (13, 175)
 (14, 164)
 (15, 163)
 (16, 164)
 (17, 161)
 (18, 158)
 (19, 161)
 (20, 156)
 (21, 154)
 (22, 159)
 (23, 144)
 (24, 153)
 (25, 150)
 (26, 140)
 (27, 141)
 (28, 141)
 (29, 139)
 (30, 138)
 (31, 134)
 (32, 134)
 (33, 135)
 (34, 128)
 (35, 127)
 (36, 120)
 (37, 121)
 (38, 120)
 (39, 117)
 (40, 121)
 (41, 114)
 (42, 109)
 (43, 109)
 (44, 109)
 (45, 103)
 (46, 109)
 (47, 103)
 (48, 99)
 (49, 99)
 (50, 94)
 (51, 94)
 (52, 89)
 (53, 87)
 (54, 91)
 (55, 87)
 (56, 84)
 (57, 81)
 (58, 81)
 (59, 73)
 (60, 81)
 (61, 77)
 (62, 71)
 (63, 74)
 (64, 110)
 (65, 111)
 (66, 121)
 (67, 121)
 (68, 116)
 (69, 107)
 (70, 93)
 (71, 111)
 (72, 98)
 (73, 99)
 (74, 94)
 (75, 90)
 (76, 86)
 (77, 73)
 (78, 76)
 (79, 77)
 (80, 69)
 (81, 94)
 (82, 78)
 (83, 78)
 (84, 76)
 (85, 74)
 (86, 71)
 (87, 80)
 (88, 72)
 (89, 78)
 (90, 71)
 (91, 72)
 (92, 69)
 (93, 71)
 (94, 67)
 (95, 67)
 (96, 59)
 (97, 64)
 (98, 66)
 (99, 63)
 (100, 63)
 (101, 65)
 (102, 61)
};
\addlegendentry{total}
\end{axis}
\end{tikzpicture}

%% file: fig/plots/HB_OH_100.tex
\begin{tikzpicture}
\begin{axis}[ 
    grid=major,
    xmin = 0, xmax = 102,
    ymin = 0, ymax = 250,
    ylabel style={},
    legend style ={ at={(0.64,0.99)},anchor=north west, draw=black,fill=white,align=left},
    xlabel={$ sec $}
]
\addplot[
   color=black,
   solid,
   thin,
   smooth
   ]
coordinates{
 (0, 198)
 (1, 188)
 (2, 198)
 (3, 176)
 (4, 198)
 (5, 182)
 (6, 181)
 (7, 190)
 (8, 180)
 (9, 173)
 (10, 175)
 (11, 177)
 (12, 174)
 (13, 170)
 (14, 156)
 (15, 163)
 (16, 160)
 (17, 163)
 (18, 146)
 (19, 171)
 (20, 155)
 (21, 159)
 (22, 152)
 (23, 145)
 (24, 140)
 (25, 147)
 (26, 141)
 (27, 135)
 (28, 147)
 (29, 131)
 (30, 137)
 (31, 136)
 (32, 128)
 (33, 132)
 (34, 127)
 (35, 119)
 (36, 116)
 (37, 130)
 (38, 114)
 (39, 117)
 (40, 121)
 (41, 113)
 (42, 112)
 (43, 100)
 (44, 113)
 (45, 108)
 (46, 102)
 (47, 98)
 (48, 103)
 (49, 97)
 (50, 96)
 (51, 93)
 (52, 89)
 (53, 90)
 (54, 87)
 (55, 86)
 (56, 82)
 (57, 84)
 (58, 79)
 (59, 76)
 (60, 77)
 (61, 74)
 (62, 68)
 (63, 71)
 (64, 67)
 (65, 66)
 (66, 65)
 (67, 62)
 (68, 59)
 (69, 57)
 (70, 56)
 (71, 54)
 (72, 49)
 (73, 49)
 (74, 49)
 (75, 47)
 (76, 43)
 (77, 41)
 (78, 40)
 (79, 38)
 (80, 34)
 (81, 35)
 (82, 32)
 (83, 30)
 (84, 25)
 (85, 28)
 (86, 24)
 (87, 22)
 (88, 19)
 (89, 18)
 (90, 16)
 (91, 14)
 (92, 11)
 (93, 10)
 (94, 8)
 (95, 6)
 (96, 4)
 (97, 1)
 (98, 0)
 (99, 0)
 (100, 0)
 (101, 0)
 (102, 0)
};
\addlegendentry{data}
\addplot[
   color=black,
   dash pattern=on 3pt off 3pt on 3pt off 3pt,
   smooth
   ]
coordinates{
 (0, 0)
 (1, 4)
 (2, 0)
 (3, 6)
 (4, 1)
 (5, 5)
 (6, 5)
 (7, 4)
 (8, 9)
 (9, 12)
 (10, 11)
 (11, 10)
 (12, 11)
 (13, 13)
 (14, 14)
 (15, 1)
 (16, 2)
 (17, 1)
 (18, 5)
 (19, 2)
 (20, 3)
 (21, 3)
 (22, 8)
 (23, 8)
 (24, 5)
 (25, 3)
 (26, 5)
 (27, 2)
 (28, 2)
 (29, 5)
 (30, 1)
 (31, 2)
 (32, 8)
 (33, 5)
 (34, 6)
 (35, 13)
 (36, 3)
 (37, 1)
 (38, 3)
 (39, 2)
 (40, 3)
 (41, 7)
 (42, 3)
 (43, 4)
 (44, 2)
 (45, 3)
 (46, 7)
 (47, 9)
 (48, 13)
 (49, 14)
 (50, 13)
 (51, 18)
 (52, 29)
 (53, 77)
 (54, 81)
 (55, 70)
 (56, 79)
 (57, 64)
 (58, 81)
 (59, 61)
 (60, 75)
 (61, 72)
 (62, 76)
 (63, 69)
 (64, 65)
 (65, 67)
 (66, 66)
 (67, 57)
 (68, 61)
 (69, 57)
 (70, 57)
 (71, 56)
 (72, 52)
 (73, 47)
 (74, 47)
 (75, 60)
 (76, 79)
 (77, 83)
 (78, 79)
 (79, 66)
 (80, 70)
 (81, 67)
 (82, 71)
 (83, 83)
 (84, 76)
 (85, 75)
 (86, 64)
 (87, 86)
 (88, 79)
 (89, 79)
 (90, 79)
 (91, 84)
 (92, 82)
 (93, 89)
 (94, 87)
 (95, 89)
 (96, 89)
 (97, 92)
 (98, 85)
 (99, 77)
 (100, 88)
 (101, 83)
 (102, 90)
};
\addlegendentry{HB\_reply}
\addplot[
   color=black,
   solid,
   smooth,
   thick
   ]
coordinates{
 (0, 198)
 (1, 192)
 (2, 198)
 (3, 182)
 (4, 199)
 (5, 187)
 (6, 186)
 (7, 194)
 (8, 189)
 (9, 185)
 (10, 186)
 (11, 187)
 (12, 185)
 (13, 183)
 (14, 170)
 (15, 164)
 (16, 162)
 (17, 164)
 (18, 151)
 (19, 173)
 (20, 158)
 (21, 162)
 (22, 160)
 (23, 153)
 (24, 145)
 (25, 150)
 (26, 146)
 (27, 137)
 (28, 149)
 (29, 136)
 (30, 138)
 (31, 138)
 (32, 136)
 (33, 137)
 (34, 133)
 (35, 132)
 (36, 119)
 (37, 131)
 (38, 117)
 (39, 119)
 (40, 124)
 (41, 120)
 (42, 115)
 (43, 104)
 (44, 115)
 (45, 111)
 (46, 109)
 (47, 107)
 (48, 116)
 (49, 111)
 (50, 109)
 (51, 111)
 (52, 118)
 (53, 167)
 (54, 168)
 (55, 156)
 (56, 161)
 (57, 148)
 (58, 160)
 (59, 137)
 (60, 152)
 (61, 146)
 (62, 144)
 (63, 140)
 (64, 132)
 (65, 133)
 (66, 131)
 (67, 119)
 (68, 120)
 (69, 114)
 (70, 113)
 (71, 110)
 (72, 101)
 (73, 96)
 (74, 96)
 (75, 107)
 (76, 122)
 (77, 124)
 (78, 119)
 (79, 104)
 (80, 104)
 (81, 102)
 (82, 103)
 (83, 113)
 (84, 101)
 (85, 103)
 (86, 88)
 (87, 108)
 (88, 98)
 (89, 97)
 (90, 95)
 (91, 98)
 (92, 93)
 (93, 99)
 (94, 95)
 (95, 95)
 (96, 93)
 (97, 93)
 (98, 85)
 (99, 77)
 (100, 88)
 (101, 83)
 (102, 90)
};
\addlegendentry{total}
\end{axis}
\end{tikzpicture}

%% file: fig/plots/packet_loss_FRIP_vs_OF.tex
\begin{tikzpicture}
\begin{axis}[ 
  grid=major,
  xmin = 0, xmax = 40,
  ymin = 0, ymax = 300,
  ylabel={Number of packets lost},
  ylabel style={},
  legend style ={ at={(0.03,0.97)},anchor=north west, draw=black,fill=white,align=left},
  xlabel={Number of demands served by the switch }
]
%\addplot[color=black, mark options={solid}, mark=square*]
\addplot[
   dash pattern=on 4pt off 1pt on 1pt off 1pt, smooth
   ]
coordinates{
  (1,2)
  (2,6)
  (3,8)
  (4,11)
  (5,16)
  (6,14)
  (7,29)
  (8,28)
  (9,36)
  (10,35)
  (11,48)
  (12,46)
  (13,52)
  (14,64)
  (15,78)
  (16,81)
  (17,84)
  (18,97)
  (19,106)
  (20,111)
  (21,123)
  (22,120)
  (23,128)
  (24,142)
  (25,151)
  (26,161)
  (27,170)
  (28,152)
  (29,173)
  (30,203)
  (31,170)
  (32,201)
  (33,193)
  (34,209)
  (35,197)
  };
\addlegendentry{OF FF (RTT 12ms)}
\addplot[
   dash pattern=on 4pt off 1pt on 1pt off 1pt on 1pt off 1pt, smooth
   ]
coordinates{
  (1,2)
  (2,4)
  (3,5)
  (4,11)
  (5,12)
  (6,15)
  (7,14)
  (8,20)
  (9,26)
  (10,28)
  (11,31)
  (12,42)
  (13,36)
  (14,44)
  (15,50)
  (16,62)
  (17,74)
  (18,75)
  (19,85)
  (20,89)
  (21,89)
  (22,100)
  (23,106)
  (24,111)
  (25,115)
  (26,127)
  (27,113)
  (28,129)
  (29,139)
  (30,134)
  (31,137)
  (32,158)
  (33,139)
  (34,208)
  (35,182)
  };
\addlegendentry{OF FF (RTT 6ms)}
%\addplot[color=black, mark options={solid}, mark=*]
\addplot[
   dash pattern=on 2pt off 1pt, smooth
   ]
coordinates{
  (1,1)
  (2,2)
  (3,6)
  (4,8)
  (5,10)
  (6,7)
  (7,11)
  (8,14)
  (9,21)
  (10,31)
  (11,28)
  (12,31)
  (13,37)
  (14,38)
  (15,50)
  (16,50)
  (17,64)
  (18,77)
  (19,76)
  (20,83)
  (21,85)
  (22,93)
  (23,90)
  (24,99)
  (25,104)
  (26,123)
  (27,128)
  (28,120)
  (29,127)
  (30,140)
  (31,136)
  (32,154)
  (33,159)
  (34,177)
  (35,183)
  };
\addlegendentry{OF FF (RTT 3ms)}
%\addplot[color=black, mark options={solid}, mark=*]
\addplot[
   dash pattern=on 1pt off 1pt,
   smooth
   ]
coordinates{
  (1,2)
  (2,2)
  (3,3)
  (4,4)
  (5,5)
  (6,9)
  (7,9)
  (8,13)
  (9,12)
  (10,18)
  (11,29)
  (12,28)
  (13,30)
  (14,35)
  (15,37)
  (16,39)
  (17,71)
  (18,71)
  (19,69)
  (20,71)
  (21,72)
  (22,79)
  (23,79)
  (24,94)
  (25,98)
  (26,113)
  (27,98)
  (28,106)
  (29,112)
  (30,118)
  (31,109)
  (32,134)
  (33,124)
  (34,160)
  (35,143)
  };
\addlegendentry{OF FF (RTT 0ms)}

\addplot[
   solid,
   smooth
   ]
coordinates{
  (1,1)
  (2,1)
  (3,2)
  (4,2)
  (5,3)
  (6,2)
  (7,3)
  (8,2)
  (9,2)
  (10,4)
  (11,5)
  (12,3)
  (13,5)
  (14,7)
  (15,5)
  (16,6)
  (17,8)
  (18,9)
  (19,13)
  (20,13)
  (21,15)
  (22,16)
  (23,20)
  (24,17)
  (25,21)
  (26,22)
  (27,23)
  (28,26)
  (29,26)
  (30,27)
  (31,29)
  (32,31)
  (33,31)
  (34,31)
  (35,30)

  };
\addlegendentry{SPIDER $\delta_7$=1ms}
\end{axis}
\end{tikzpicture}